\documentclass[longnamesfirst,apjl]{emulateapj}
\usepackage{apjfonts,natbib,epsfig}
\tighten

\newcommand{\beq}{\begin{equation}}
\newcommand{\eeq}{\end{equation}}
\newcommand{\beqa}{\begin{eqnarray}}
\newcommand{\eeqa}{\end{eqnarray}}

\begin{document}

\title{
Dissipationless Merging and the Assembly of Central Galaxies
}

\author{Asantha Cooray$^1$ and Milo\v s Milosavljevi\'c$^2$}
\affil{$^1$Department of Physics and Astronomy, University of California, Irvine, CA 92617\\
$^2$Theoretical Astrophysics, California Institute of Technology, Pasadena, CA 91125}
\righthead{ASSEMBLY OF CENTRAL GALAXIES}
\lefthead{COORAY \& MILOSAVLJEVI\'C}
\begin{abstract}
We reanalyze the galaxy-mass correlation function 
measured by the Sloan Digital Sky Survey to obtain host dark matter halo masses 
at galaxy and galaxy group scales.
We extend the data to galaxy clusters 
in the 2MASS catalog and study the relation between central galaxy luminosity and halo mass. 
While the central galaxy luminosity scales as $\sim M^{0.7-0.8}$ at low masses,
the relation flattens to  $\sim M^{<0.3}$ above $\sim 4\times 10^{13}M_\odot$.
The total luminosity of galaxies in the halo, however, continues to grow
as a power-law $\sim M^{0.8-0.9}$. 
Starting from the hypothesis that the central galaxies grow by merging (``galactic cannibalism''), we develop a simple model for the evolution of their luminosities as a consequence of the accretion of satellite galaxies.  The luminosity-mass relation flattens when the time scale on which dynamical friction induces orbital decay in the satellite galaxies exceeds the age of the dark matter halo.  Then, the growth of the central galaxy is suppressed as it can cannibalize only the rare, massive satellite galaxies.  The model takes the dependence of the total luminosity of galaxies in a halo on its mass and the global galaxy luminosity function as input, and reproduces the observed central galaxy luminosity-mass relation over three decades in halo mass, $(10^{12}-10^{15})M_\odot$.
 The success of the model suggests that gas cooling and subsequent star formation did not play an important role in the final assembly of central galaxies from sub-$L_\star$ precursors.


\end{abstract}

\section{Introduction}
\label{sec:introduction}

Our understanding of galaxy formation and evolution is still incomplete. 
The mass function of dark matter halos is routinely measured in numerical simulations.
If a linear relation is assumed between the galaxy luminosity and the halo mass, 
the abundance of galaxies at the faint and the bright ends of the luminosity range is significantly below the expected \citep{Vale:04,van:04}.
The standard picture of galaxy formation (e.g., \citealt{Rees:77,White:78}) 
postulates that as dark matter halos virialize, the gas inside the halos is heated to the virial
temperature and then cools to form galaxies. 
While the dearth of faint dwarf galaxies can be explained in semianalytic models
by invoking a  variety of feedback mechanisms that expel gas from 
small dark matter halos at early times,  these mechanisms eventually yield an overabundance of
bright galaxies as the expelled gas eventually cools in massive halos 
to form luminous galaxies (e.g., \citealt{Benson:03}).

Motivated by the discovery that the temperature distribution of gas in virializing halos in numerical simulations is bimodal \citep{Katz:03}, 
and by the lack of evidence for significant cooling flows in galaxy clusters (e.g., \citealt{Fabian:01}),
\citet{Binney:04} suggested that only 
the colder component cools to form galaxies, while the hotter component remains at the virial temperature. 
 One mass scale characterizing galaxy formation would then be that
below which the shocks cannot be sustained, the gas is not heated, and all the gas in the colder component streams to the halo center and forms a galaxy. Models of shock formation
suggest this critical mass is small, $M_{\rm shock}\sim (1-6) \times 10^{11}M_\sun$ \citep{Dekel:04}.  A similar result is derived in \citet{Maller:04} based on models of multi-phase galaxy formation.

Since galaxies do not continue to grow by accreting hot gas, their stellar masses can increase only by
accreting other galaxies. Thus, we expect that the galaxy growth on scales above $M_{\rm shock}$ is simply governed by the physics of dissipationless merging of pre-existing galaxies.  The merging of galaxies tracks
the hierarchical assembly of the host dark matter halos \citep{Ostriker:75,Ostriker:77,Merritt:85,Dubinski:98}.
Here, a second characteristic scale is expected, which is associated with the
efficiency by which galaxies accrete other galaxies.
Above this second scale, smaller galaxies that have merged into a larger halo do not have enough time to reach the halo center and aggregate into one final central galaxy.  

 In \S~\ref{sec:correlation}, we extract the dependence of central galaxy luminosity on halo mass in galaxies imaged with SDSS and 2MASS surveys. We provide evidence for a characteristic scale in this dependence at $\sim (1-6)\times10^{13} M_\sun$.
In \S~\ref{sec:derivation}, we derive a relation between the luminosity of 
the central galaxy and the mass of its host dark matter halo from first principles.
The derived relation fits the data from galaxy scales
$\sim 10^{12} M_{\sun}$ to cluster scales
$\sim 10^{15}M_{\sun}$.   

We adopt the current concordance cosmological model consistent with WMAP \citep{Spergel:03}. Unless otherwise noted, $M$ refers to the virial mass of a halo 
out to an overdensity of $200$, and $r_{\rm vir}$ is the associated virial radius.

\section{The Galaxy-Dark Matter Correlation Function}
\label{sec:correlation}

The luminosities of central galaxies in clusters and groups can be extracted directly from IR imaging data 
in the 2MASS survey, while the masses can be inferred form the X-ray temperatures \citep{LinMohr:04}.
As a direct mass measurement of the halo mass is
impossible at galaxy scales,
we extract the masses from tangential 
shear measurements in weak gravitational lensing around a sample of foreground galaxies 
(e.g., \citealt{Brainerd:96,Fischer:00}). 
The measurements from the Sloan Digital Sky Survey (SDSS; \citealt{York:00}), 
binned as a function of the galaxy luminosity \citep{McKay:01}, 
allow one to extract the masses of the host halos 
of the foreground galaxies, similar to \citet{Guzik:02} and \citet{Yang:03}. 

In galaxy-galaxy lensing, one measures tangential shear around foreground galaxies as a function
of the project distance $R$. The tangential shear is the ratio of the excess surface density to the critical surface density, $\gamma_{\rm t}(R) \equiv \Delta \Sigma(R)/\Sigma_{\rm crit}$. The critical density equals $\Sigma_{\rm crit}= c^2 D_{\rm s}/4 \pi G D_{\rm l} D_{\rm ls}$, where $D_{\rm s}$ is the distance to the source, $D_{\rm l}$ is the distance to the lens, and $D_{\rm ls}$ is the distance from the lens to the source.  The excess is defined as the difference between the local and the average surface density, $\Delta\Sigma(r)\equiv \Sigma(r)-\bar \Sigma$, where $\bar{\Sigma}(R) = 2R^{-2}\int_0^{R}\Sigma(R') R' dR'$ is the mean surface density within $R$.

The surface density 
 can be expressed in terms of the galaxy-mass correlation function, $\xi_{\rm gm}(r)$, via $\Sigma(R) = \bar{\rho} \int
\xi_{\rm gm} \sqrt{R^2+z^2}dz$. 
The correlation function between galaxies and dark matter can be related to the cross power spectrum
between the two through a Fourier transform, $\xi(r) = (2\pi)^{-3} \int P_{\rm gm} e^{i {\bf k} \cdot {\bf r}}d^3 k $. Within the halo paradigm \citep{Cooray:02}, the
cross power spectrum can be constructed from the halo
mass function $dn/dM$ \citep{Jenkins:01}, the density profile of a halo $\rho(r,M)$, and information on how the halos are distributed with respect to the linear density fluctuations.  We assume that the density profile is given by
the NFW function (\citealt{Navarro:97}; but see \citealt{Merritt:05}).
The concentration parameter of the NFW profile is a function of the halo mass \citep{Bullock:01}. 

To characterize multiple galaxies sharing the same halo, we specify number of galaxies of a given luminosity in a halo of a given mass, which is the conditional halo occupation number. We employ
 the most recent description of the halo occupation distribution \citep{Kravtsov:04}. We distinguish 
between central and satellite galaxies.  The satellites galaxies are described by a
mean occupation, $\langle N_{\rm sat}\rangle = [(M-M_0)/M_1]^\lambda$,
where the parameters $M_0\sim(10^{12}-10^{13})M_\odot$, $M_1\sim(10^{13}-10^{14})M_\odot$, and $\lambda\sim 1$, are given in Table 1 of \citet{Zheng:04}.  On the scales of interest, the occupation number of central galaxies is unity, $\langle N_{\rm cen}\rangle\sim1$.

The contributions to the cross power spectrum from central and satellite galaxies are,
\begin{eqnarray}
P_{\rm gm,cen}(k) &=& \frac{1}{\bar n_{\rm c}}\int \frac{M}{\bar\rho} \frac{dn}{dM} \langle N_{\rm cen} \rangle  u(k,M) dM , \nonumber \\
P_{\rm gm,sat}(k) &=& \frac{1}{\bar n_{\rm s}} \int\frac{M}{\bar\rho}  \frac{dn}{dM} \langle N_{\rm sat} \rangle  u(k,M) u_{\rm g}(k,M) dM \, .
\end{eqnarray}
Here, $u(k,M)$ and $u_{\rm g}(k,M)$ are the normalized Fourier transforms of the halo and galaxy density profiles, respectively. For simplicity, we set $u_{\rm g}(k,M) = u(k,M)$. 
We also define the average density of mass, $\bar{\rho} = \int (dn/dM) M dM$, 
and galaxies, $\bar{n}_i = \int\langle N_i \rangle (dn/dM) dM$, where $i$ distinguishes
central and satellite galaxies. We have ignored 
the large scale clustering between halos 
(e.g., there can be a correlation between galaxies in one halo and dark matter in another). 
This term is not a significant contributor to
the galaxy-mass correlation at the radii of interest ($< 1h^{-1}$ Mpc) as it 
changes the mass estimate by $\lesssim 1\%$ \citep{Guzik:02,Mandelbaum:04}.

\begin{figure*}[!t]
\centerline{\psfig{file=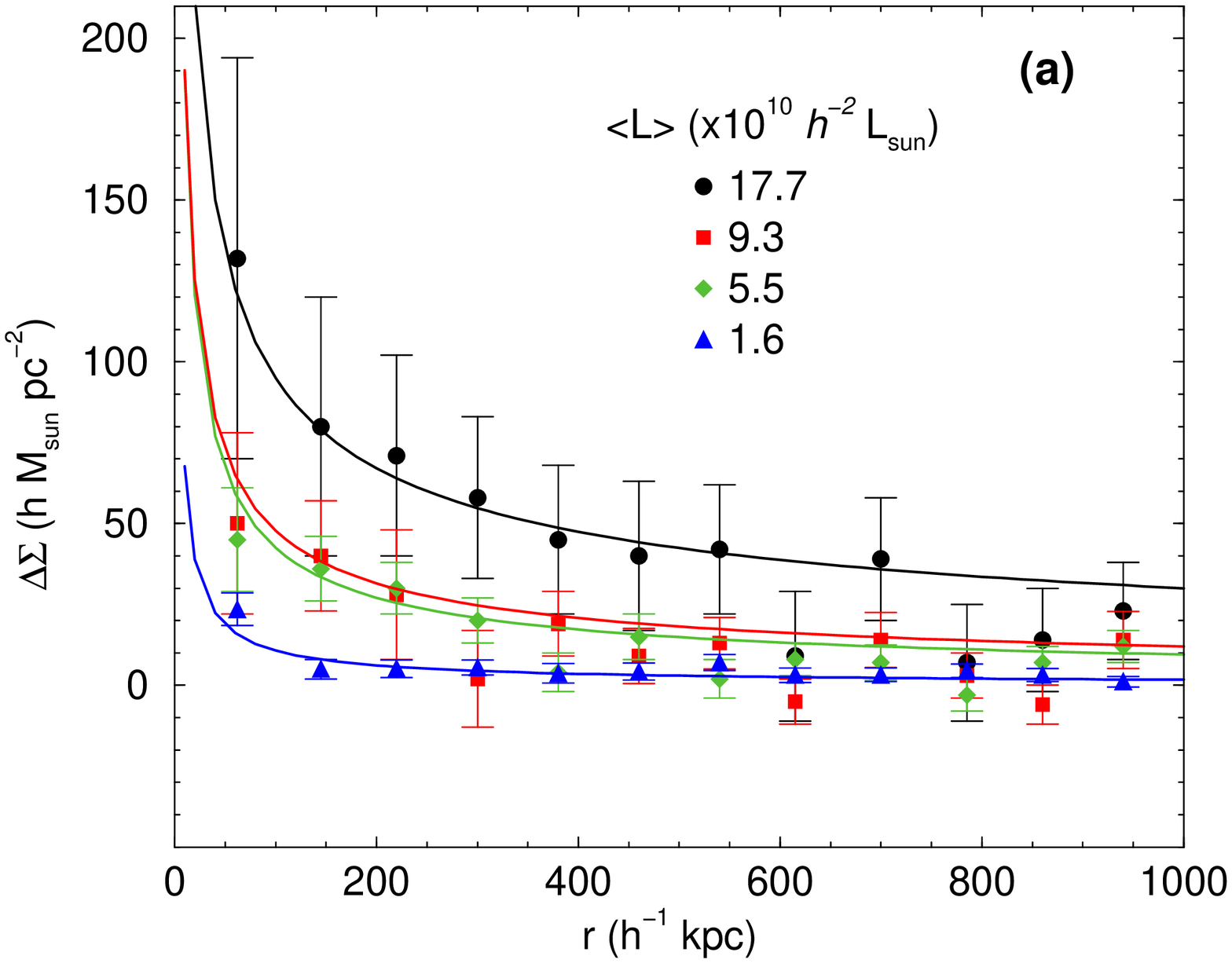,width=3.0in,angle=-90}
\psfig{file=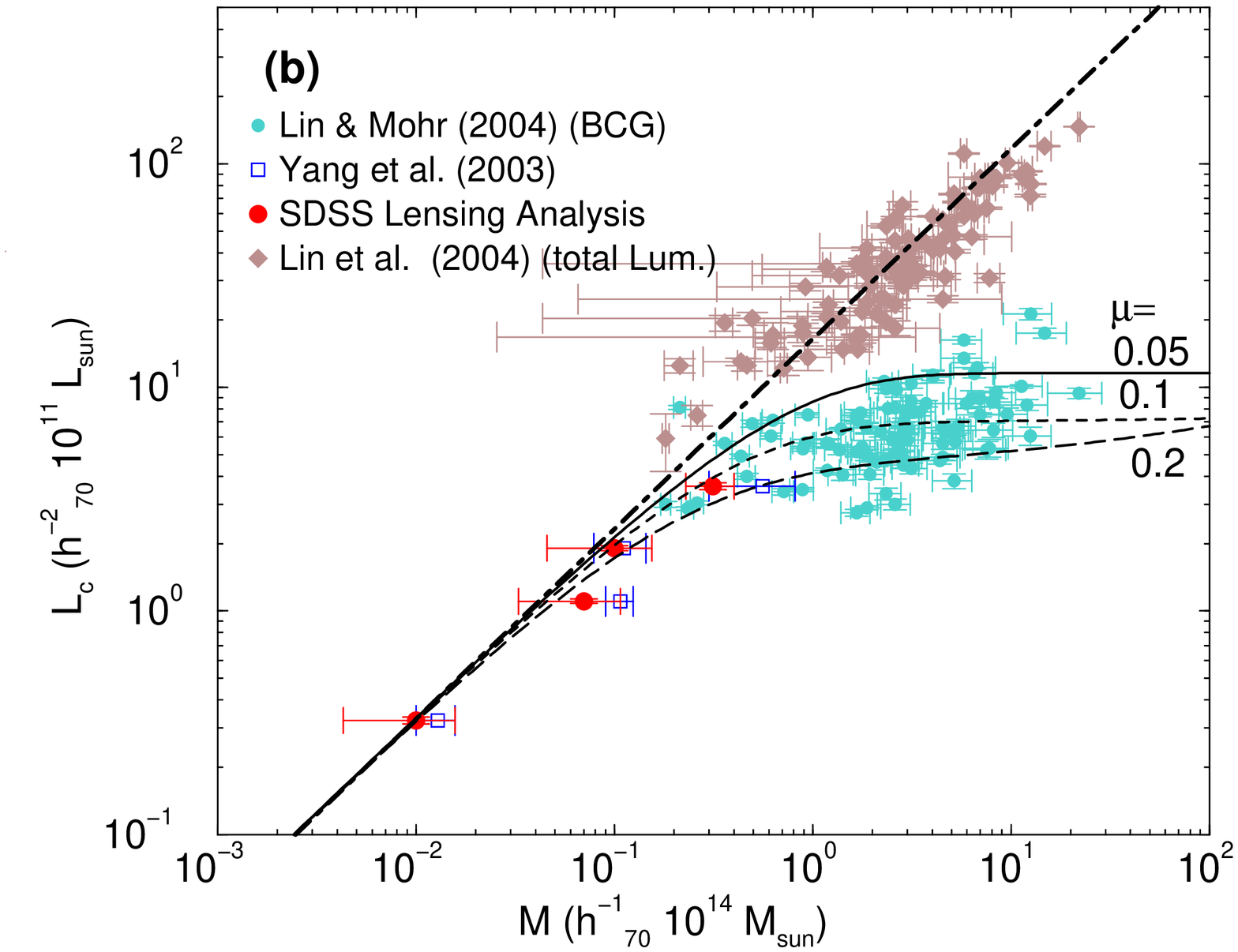,width=3.0in,angle=-90}}
\caption{(a) The galaxy-mass correlation function, expressed as the
excess surface density $\Delta \Sigma(R)$ as a function of the projected 
radius $R$. We show the SDSS $z'$-band data binned into four luminosity bins 
(from \citealt{McKay:01}), and fits based on the halo model, as described in \S~\ref{sec:correlation}. The model fitting procedure
allows us to estimate the average halo mass in which 
the foreground galaxies are contained in each of the
the four luminosity bins.  (b) Central galaxy luminosity as a function of the
halo mass. The data points at the low mass end ({\it large circles}) 
were determined by model fits to the
SDSS galaxy-mass correlation function (see \S~\ref{sec:correlation}). 
For comparison, 
we also show the masses estimated by \citet{Yang:03} using the same galaxy-mass correlation function ({\it squares}),
based on direct NFW profile fitting to the data or 
on the scaling of the mass within 260 kpc,
as determined by \citet{McKay:01}, to the virial radius obtained 
from numerical simulations.
At the high mass end, we show a direct measurement 
of galaxy luminosity and halo mass for a sample of galaxy groups and clusters from
Lin \& Mohr (2004; {\it small circles}). 
We also show the total luminosity of galaxy groups and clusters based on
data of Lin et al.\ (2004; {\it diamonds}). The dot-dashed line is a power law fit to the combined lensing and Lin et al.\ data with logarithmic slope $\beta=0.85$. The curves, from top to bottom, our models for the central galaxy luminosity-halo mass relation with the parameter $\mu=(0.05,0.1,0.2)$, respectively, as discussed in \S~\ref{sec:derivation}.}
\label{fig:sloan}
\end{figure*}

In Figure \ref{fig:sloan}a, we show the \citet{McKay:01} measurement of excess surface density, separated into four
$z'$-band luminosity bins.  Overlaid are the best fit models for $\Delta\Sigma(R)$ calculated as described above.
Given that the concentration parameter is a function of mass, the fitting is done over a single parameter, the halo mass.
We concentrate on the $z'$-band data because
the relation we expect to derive between the central galaxy luminosity
and the halo mass is studied at galaxy cluster scales in the $K$-band.  

In Figure \ref{fig:sloan}b, we compile our mass estimates, as well as the data from \citet{LinMohr:04} for
central galaxy luminosity in groups and clusters, and the total luminosities in these systems \citep{Lin:04}. 
Logarithmic slope of the relation between the halo mass and the central galaxy luminosity is different at low and high mass scales.  At the low mass end, the fits 
to the galaxy-galaxy lensing data suggest $L_{\rm c} \propto M^{0.7-0.8}$, consistent with \citet{Guzik:02} and \citet{Yang:03}.  
At the high mass end, the relation is shallower, $L_{\rm c} \propto M^{0.25}$ \citep{LinMohr:04}.
Such a relation is also required when interpreting the statistics of the 2dF galaxy group catalog \citep{Yang:05}.

The relation between the halo mass and the {\it total} luminosity of galaxies in the halo, however, 
remains a power-law over three decades in mass  $L_{\rm tot} \propto
M^{\beta}$ with $\beta\sim 0.85$.  Such a power-law is consistent with the dependence of the {\it stellar} mass-to-light ratio on mass, $M/L \sim
M^{1/6}$ \citep{Bender:92}. At the high-mass end, the power-law could also be related to the mass function of subhalos
within a halo of mass $M$. 
With the scaling $N_{\rm s}(m|M) dm \sim m^a M^{a-1} dm$ for the number of
 subhalos with masses between $m$ and $m+dm$ \citep{Vale:04}, the total luminosity of the parent halo scales
as $L_{\rm tot}(M) \approx \int N_{\rm s}(m|M) L_{\rm s}(m) dm \propto M^{a-1}$, 
where $L_{\rm s}(m)$ luminosity of the galaxy in the subhalo, which vanishes in subhalos too small to contain galaxies.
The slope of the $L_{\rm tot}$--$M$ relation is then
consistent with the numerical estimates
$a\approx1.8-1.9$ \citep{DeLucia:04}.

\section{Central Galaxy Luminosity-Halo Mass Relation}
\label{sec:derivation}

We now calculate the rate at which the luminosity of the central galaxy grows due to the accretion of satellite galaxies in the same dark matter halo.  The accretion is moderated by the dynamical friction time scale, $t_{\rm df}$, on which a satellite galaxy  sinks in the potential of the primary halo.  
The effective total mass of a satellite galaxy is augmented by the bound dark matter mass left behind from the epoch when the satellite was at the center of an isolated halo.  The central galaxy luminosity thus grows at the rate 
\beq
\label{eq:dldt}
\frac{dL}{dt}=\int_0^{L} \frac{dN}{dL_{\rm s}} \frac{L_{\rm s}}{t_{\rm df}(L_{\rm s})} dL_{\rm s} ,
\eeq
where $N(L_{\rm s})$ is the number of satellite galaxies with luminosities less than $L_{\rm s}$ (the integrated luminosity function).  

A satellite halo experiences torque $T_{\rm df}=|{\bf r}\times{\bf F}_{\rm df}|$, where ${\bf F}_{\rm df}$ is the force of dynamical friction, $|{\bf F}_{\rm df}|= 4\pi f_{\rm df } G^2 M_{\rm s}(r_{\rm s})^2 \rho(r)\ln(\Lambda) / v_{\rm s}^2$ (\citealt{Chandrasekhar:43}; but see \citealt{Tremaine:84}). For circular orbits, the velocity of the satellite is the circular velocity in the primary halo $v_{\rm s}=\sqrt{GM(r)/r}$; $M(r)$ is the mass of the primary halo contained within this radius; $\rho(r)$ is the density; $M_{\rm s}(r_{\rm s})$ is the mass of the satellite contained within the radius $r_{\rm s}$, measured from the center of the satellite, at which the satellite is tidally truncated, and $\ln\Lambda$ is the Coulomb logarithm.
In numerical simulations of satellites in halos \citep{Velazquez:99,Fellhauer:00}, $\ln\Lambda\approx 2$; for an explanation of small $\ln\Lambda$ see Appendix A of \citet{Milosavljevic:01}.  The numerical factor $f_{\rm df}$ is of order unity and depends on the orbit of the satellite and on the orbital phase space distribution of dark matter.  The satellite is tidally truncated at the radius where its average density equals that of the host halo within the orbit of the satellite, $M_{\rm s}(r_{\rm s})/r_{\rm s}^3$.  

Let $J_{\rm s}=\sqrt{GM(r)r}$ be the specific angular momentum of the satellite.  The satellite spirals toward the center of the primary halo on a time scale $t_{\rm df}=(M_{\rm s}/T_{\rm df})(dJ_{\rm s}/d\ln r)$.  In the outer parts of halos described by the NFW profile, $d\ln J_{\rm s}/d\ln r\approx 0.7-1$. The time scale is the longest when the satellite has just entered the virial radius of the primary halo and thus we evaluate $t_{\rm df}$ at $r=r_{\rm vir}$; the average densities of the two halos when they are touching at the virial radii, $\bar\rho$, are both equal to $200$ times the universal matter density, while the density at radius $r_{\rm vir}$ is a factor of $\sim(4-7)$ {\it smaller} than the average density, depending on the halo concentration. We denote the ratio $\bar\rho/\rho(r_{\rm vir})\equiv\eta\approx 5$.  If the orbit of the satellite is eccentric when it first enters the primary (a very likely circumstance!), the effective density that gives rise to the dynamical drag is {\it larger} than at the edge of the halo.  We do not explore these complications and subsume the dependence on the halo orbit in the factor $f_{\rm df}$.  The dynamical friction then equals
\beq
\label{eq:df}
t_{\rm df}=\frac{1}{2\sqrt{3\pi}}\frac{\eta}{f_{\rm df}\ln\Lambda}\frac{1}{\sqrt{G\bar\rho}}\frac{M}{M_{\rm s}} .
\eeq
Therefore, $t_{\rm df}$ exceeds the dynamical time of the primary halo $\sim (G\bar\rho)^{-1/2}$ by a factor proportional to the mass ratio of the two halos.

The luminosity function of galaxies sharing the same dark matter halo can be modeled by
\beq
\label{eq:lf}
\frac{dN}{dL}(L)=\Phi_\star\left(\frac{L}{L_\star}\right)^\alpha e^{-L/L_\star} ,
\eeq
where $L_\star\approx 8.3\times10^{10} L_\sun$ is a characteristic luminosity scale, $-1.3\lesssim\alpha\lesssim-0.8$ (\citealt{Lin:04}, and references therein; $\alpha\lesssim-1$ is appropriate for field galaxies and $\alpha\approx -0.8$ for clusters), and $\Phi_\star$ is a normalization factor 
set by the requirement that the total luminosity in the cluster equals $L_{\rm tot}(M)$. Thus, 
$\Phi_\star=L_{\rm tot}(M)/L_\star^2\Gamma(2+\alpha)$. 
 Substituting equations (\ref{eq:df}) and (\ref{eq:lf}) into equation (\ref{eq:dldt}) we obtain
\beqa
\label{eq:dldt_expanded}
\frac{dL}{dt}&=&\frac{2\sqrt{3\pi}f_{\rm df}\ln\Lambda}{\Gamma(2+\alpha) \eta}\frac{L_{\rm tot}(M)}{L_\star^2}\frac{\sqrt{G\bar\rho}}{M}\nonumber\\& &\times\int_{L(M_{\rm min})}^L\left(\frac{L_{\rm s}}{L_\star}\right)^\alpha e^{-L_{\rm s}/L_\star} M_{\rm s}(L_{\rm s}) L_{\rm s}dL_{\rm s} ,
\eeqa
where $M_{\rm s}(L_{\rm s})$ is the parent dark matter mass associated with the satellite of luminosity $L_{\rm s}$, and $L(M_{\rm min})\leq L(M)$ is the luminosity associated with the smallest satellite, having mass $M_{\rm min}(M)$, that can 
be accreted to the halo center in a growth time of the primary halo, $1/\tau H_0$, where $\tau\sim 1$ \citep{Wechsler:02}.  
The minimum mass can be then be obtained from the condition $t_{\rm df}(M_{\rm s}=M_{\rm min})=1/\tau H_0$.  We divide equation (\ref{eq:dldt_expanded}) by $dM/dt=\tau M H_0$ to obtain the rate of increase in luminosity of the central galaxy per increase in the halo mass as
\beqa
\label{eq:dldm}
\frac{dL}{dM}&=&\frac{30f_{\rm df}\ln\Lambda}{\Gamma(2+\alpha)\eta\tau}\frac{L_{\rm tot}(M)}{L_\star^2} \frac{\sqrt{\Omega_{\rm m}}}{M^2} \nonumber\\& &\times\int_{L(M_{\rm min})}^L\left(\frac{L_{\rm s}}{L_\star}\right)^\alpha e^{-L_{\rm s}/L_\star}  M_{\rm s}(L_{\rm s}) L_{\rm s} dL_{\rm s} , 
\eeqa
where we have used the virial density relation to critical density 
$\bar\rho = 200\times 3\Omega_{\rm m} H_0^2/8\pi G$, where $\Omega_{\rm m}\approx0.27$.

To obtain a rough idea about the meaning of equation (\ref{eq:dldm}), we note that we are interested in the $L$--$M$ relation for galaxies brighter than $L_\star$.  Then the integral can be approximated by the integrand evaluated at the lower limit and multiplied by $L_\star$.  Since $M_{\rm min}=\mu M$ with $\mu\equiv \eta\tau /30\sqrt{\Omega_{\rm m}}f_{\rm df}\ln\Lambda$,
\beq
\label{eq:dldm_limit}
\frac{dL}{dM}\approx \frac{1}{\Gamma(2+\alpha)} \frac{L_{\rm tot}(M)}{M}
\left[\frac{L(\mu M)}{L_\star}\right]^{1+\alpha} e^{-L(\mu M)/L_\star} .
\label{dldm}
\eeq

\section{Results and Discussion}
\label{sec:discussion}

We integrate the delay differential equation (\ref{eq:dldm_limit}) numerically from small to large masses, where at the small masses we assume that $L(M)=L_{\rm tot}(M)$, where $L_{\rm tot}(M) = L_\star (M/M_\star)^\beta$ with $M_\star\approx 3.5\times10^{12} M_\sun$. In the regime $L(\mu M)\ll L_\star$ where the exponential factor can be neglected, we obtain $L(M)\sim L_\star (M/M_\star)^{-\beta/\alpha}$, i.e., the luminosity of the central galaxy is a power law, similar to that describing the total luminosity of galaxies in the halo.  On the scale $M_{\rm crit}\sim M_\star/\mu$ the power-law behavior breaks, and the growth of the luminosity is suppressed.  

Using $\alpha=-1$, $\beta=0.85$, $\tau=1$, $\eta=5$, $\Omega_{\rm m}=0.3$, and $\ln\Lambda=2$, we estimate $\mu\approx 0.16$.  As discussed above, the evaluation of the exact value of $\mu$ is beyond the scope of this analysis and may vary from one halo to another.  We compare the predictions of the model by plotting $L(M)$ obtained by direct integration of equation (\ref{eq:dldm_limit}) over the data in Figure \ref{fig:sloan}b.  Because of the uncertainty in the precise value of $\mu$, we present curves for $\mu=(0.05,0.1,0.2)$.
For $\mu=0.1$, the critical mass scale of the luminosity growth retardation is
$M_{\rm crit}\approx 3\times10^{13} M_\sun$, which is naturally
associated with the halos in which the time scale on which the satellites merge with the central galaxy is equal to the age of the primary halo.
This is an additional fundamental scale characterizing
galaxy formation. This 
scale is unlikely to be directly observed in the 
luminosity function as the statistical averaging over several decades in halo mass, implicit in the evaluation of the function, erases its signatures.

As evident in Figure \ref{fig:sloan}b, $M_{\rm crit}$, identified with the break in the $L$--$M$ relation, is to some degree sensitive to the value of $\mu$, which is in turn sensitive to the precise calibration of the dynamical friction force, as well as to the orbit of the satellite at the point of initial entry into the primary halo.  The value of $\mu$ is also sensitive to $\tau$, the dynamical age of the primary halo.  Both uncertainties plausibly give rise to a factor of 2 variation in $\mu$.  This can explain the large scatter in the observed central galaxy luminosity-halo mass relation
in luminous groups and clusters. 
Except for a few outliers, 
the derived relation reproduces the data over three decades in mass. 

The success in deriving the
relation from first principles and without reliance on detailed numerical simulations or semi-analytical models encourages us to propose several general conclusions. In semi-analytic models of galaxy formation that are based on efficient gas cooling,
the high mass end of the luminosity function is 
generally overabundant (``the overcooling problem;'' e.g., \citealt{Benson:03}). The same high-end of the luminosity function is dominated by
central galaxies above $L_\star$.  Given that we were able to describe
the luminosity growth of such galaxies through dissipationless merging, it is unlikely that additional cooling inside groups and clusters contributes significantly to their stellar mass budget. Thus, gas cooling must be generally suppressed, either through feedback or heating, over a wide range of
mass scales ranging from clusters down to $L_\star$ galaxies (see also \citealt{Maller:04}). 
While the exact mechanism of cooling suppression
remains a mystery, we believe that it is not restricted to 
clusters with temperature above $\sim1\textrm{ keV}$, as suggested by \citet{Fabian:04}.

Our results support the paradigm for the formation of giant galaxies with two fundamental scales, namely, that related to the efficiency of shock-heating $\sim(1-6) \times 10^{11}M_{\sun}$ \citep{Dekel:04}, and that related to the efficiency of merging $\sim(1-6)\times 10^{13}M_{\sun}$. The standard $M_\star$ galaxies, which correspond to
$L_\star$ galaxies on the luminosity scale, belong between these scales. Since $M_{\rm crit}$, interpreted as the scale at which the $L$--$M$ relation exhibits a break, is a consequence of the drop in the merging efficiency, it also may be possible to explain $M_\star$ and $L_\star$ on the basis of dissipationless merging.

\acknowledgements
A.~C.\ thanks the participants of the workshop 
{\it The Future of Cosmology with Clusters of Galaxies} and members of Cosmology groups at Caltech and U.~C.\ Irvine
for useful discussions.
M.~M.\ thanks A.~Berlind and D.~Weinberg for inspiring discussions. 
M.~M.\ was supported at Caltech by a postdoctoral fellowship from the Sherman Fairchild Foundation.

\end{document}